\documentclass[11]{article}
\usepackage[utf8]{inputenc}

\usepackage{jcappub}
\usepackage{color}
\usepackage{amsfonts, amssymb,amsmath,bm}
\usepackage{slashed}
\usepackage{braket}
\usepackage[normalem]{ulem}
\usepackage{comment}
\usepackage{xspace}
\usepackage{multirow}
\usepackage{subcaption}
\usepackage{placeins}
\usepackage{afterpage}
\usepackage{booktabs}
\usepackage{float}
\usepackage{bbold}

\newcommand{\kv}{\mathbf{k}}
\newcommand{\nv}{\mathbf{n}}

\newcommand{\qv}{\mathbf{q}}
\newcommand{\pv}{\mathbf{p}}

\newcommand{\cov}{\mathbf{C}}
\newcommand{\de}{\mathrm{d}}
\newcommand{\dv}{\mathbf{d}}
\newcommand{\nuq}{\nu_{\qv_{12}}}
\newcommand{\del}{\delta}

\usepackage[T1]{fontenc}
\DeclareMathSizes{12}{20}{4}{10}

\newcommand{\lan}{\langle}
\newcommand{\ran}{\rangle}

\newcommand{\be}{\begin{equation}}
\newcommand{\ee}{\end{equation}}
\newcommand{\bea}{\begin{eqnarray}}
\newcommand{\eea}{\end{eqnarray}}
\newcommand{\bdm}{\begin{displaymath}}
\newcommand{\edm}{\end{displaymath}}

\newcommand{\desi}{DESI}

\newcommand{\tot}{{\rm tot}}

\definecolor{ForestGreen}{rgb}{0.13, 0.55, 0.13}
\definecolor{airforceblue}{rgb}{0.36, 0.54, 0.66}
\definecolor{orange}{rgb}{1.0, 0.5, 0.0}
\definecolor{alizarin}{rgb}{0.82, 0.1, 0.26}
\definecolor{brilliantlavender}{rgb}{0.96, 0.73, 1.0}

\newcommand{\kMpc}{\, h \, {\rm Mpc}^{-1}}

\title{Bispectrum non-Gaussian Covariance in Redshift Space}

\author[a,b,c,d]{Jacopo Salvalaggio,}
\author[e]{Lina Castiblanco,}
\author[f]{Jorge Nore\~ na,}
\author[b,c,d]{Emiliano Sefusatti,}
\author[a,b,c,d]{Pierluigi Monaco}

\affiliation[a]{Dipartimento di Fisica, Sezione di Astronomia, Universit\`a di Trieste, via Valerio 2, 34127 Trieste, Italy}
\affiliation[b]{Istituto Nazionale di Astrofisica, Osservatorio Astronomico di Trieste, via Tiepolo 11, 34143 Trieste, Italy}
\affiliation[c]{Institute for Fundamental Physics of the Universe, Via Beirut 2, 34151 Trieste, Italy}
\affiliation[d]{Istituto Nazionale di Fisica Nucleare, Sezione di Trieste,  via  Valerio  2,  34127 Trieste,  Italy}
\affiliation[e]{School of Mathematics, Statistics and Physics, Newcastle University, Herschel Building, NE1 7RU Newcastle-upon-Tyne, U.K.}
\affiliation[f]{Instituto de F\'isica, Pontificia Universidad Cat\'olica de Valpara\'iso, Casilla 4950, Valpara\'iso, Chile}
\emailAdd{jacopo.salvalaggio@inaf.it, lina.castiblanco-tolosa@ncl.ac.uk, jorge.norena@pucv.cl, emiliano.sefusatti@inaf.it, pierluigi.monaco@inaf.it}

\abstract{
We provide an analytical description of the galaxy
bispectrum covariance and the power spectrum-bispectrum cross-covariance in redshift space that captures the dominant non-Gaussian contributions.
The Gaussian prediction for the variance of the halo bispectrum monopole significantly underestimates numerical estimates particularly for squeezed triangles, that is bispectrum triangular configurations where one side is much smaller than the other two, whereas the effect is relatively less important when considering the quadrupole.
We propose an expression for the missing non-Gaussian contribution valid in the squeezed limit that requires an accurate modeling of the bispectrum alone.
We validate our model against the numerical covariance estimated from a large suite of mock catalogs and find that it accurately predicts the variance as well as the dominant off-diagonal terms.
We also present an expression for the cross-covariance between power spectrum and bispectrum multipoles and likewise find it to provide a good description of the numerical results.
}

\begin{document}

\maketitle

\section{Introduction}

The analysis of the the galaxy bispectrum has received significant attention in the last years, slowly becoming a standard component in a complete cosmological study of spectroscopic galaxy redshift surveys \cite{GilMarinEtal2017, PhilcoxIvanov2022, DAmicoEtal2022B, CabassEtal2022, CabassEtal2022B, IvanovEtal2023}.

Measurements of the galaxy bispectrum multipoles in redshift space are characterized by a signal distributed over a large number of triangular configurations (see, e.g. \cite{SefusattiEtal2006, ChanBlot2017, HahnEtal2020, HahnVillaescusaNavarro2021}). This implies that a standard analysis requires the accurate evaluation of a rather large covariance matrix for the data-vector including power spectrum and bispectrum multipoles. This has been obtained in the past from a large set of mock catalogs (see, e.g. \cite{KitauraEtal2016} for the BOSS survey) but a straight-forward numerical estimate of covariance properties is becoming more challenging for current and up-coming surveys such as, for instance, {\em Euclid} \cite{LaureijsEtal2011}, or \desi~ \cite{AghamousaEtal2016}, since the large number of realizations required is limited by the high-resolution necessary for a proper description of the galaxy samples. This led to several proposals for data compression (see \cite{FergussonReganShellard2012, ByunEtal2017, GualdiEtal2018, GualdiEtal2019B, PhilcoxEtal2021, ByunEtal2021, ByunKrause2022} for recent applications to the bispectrum problem) or methods to reduce the sampling noise typically affecting numerical estimates of the covariance matrix \cite{DodelsonSchneider2013, PercivalEtal2014, HartlapSimonSchneider2007,SellentinHeavens2016}.   

Another possibility is given by an analytical prediction of the covariance properties of the observable at hand. As an example, the case of the galaxy power spectrum multipoles, including all non-Gaussian contributions and the BOSS survey geometry effects, has been studied in \cite{WadekarScoccimarro2020} (but see also \cite{MohammedSeljakVlah2017, SugiyamaEtal2020, TaruyaMishimichiJeong2021}) and applied to data analysis in \cite{WadekarIvanovScoccimarro2020}, while the configuration-space 2-point function has also been the topic of several works \cite{GriebEtal2016, PhilcoxEisenstein2019, HouEtal2022}. 

The case of the galaxy bispectrum has also been considered in a few instances in the past. A comparison between the Gaussian theoretical prediction and numerical results can be found, in real space and for the redshift-space monopole in \cite{ColavincenzoEtal2019}. In both cases the analytical model is shown to underestimate the numerical estimate, based on 300 full N-body simulations of a dark matter halo distribution. The study in \cite{RizzoEtal2023} provides a second, more detailed comparison of the Gaussian model, including higher-order multipoles and accounting for discreteness effects in mode counting based on a much more precise determination of the numerical covariance matrix obtained from 10,000 catalogs produced with the approximate method implemented in the \texttt{Pinocchio} code \cite{MonacoTheunsTaffoni2002, MunariEtal2017}. These works show that the Gaussian prediction fails (up to 20\%) in particular for squeezed triangular configurations, where one wavenumber is much smaller than the other two. The relevance of non-Gaussian contributions for these specific triangles has been pointed-out already in \cite{Barreira2019} and explored extensively, in real space by \cite{BiagettiEtal2022}, taking advantage of a large set of more than 2,000 N-body simulations. This work proposes a simple, approximate modeling of such non-Gaussian contributions, that can account for most of the off-diagonal bispectrum covariance elements. The implications for the determination of primordial non-Gaussian parameters, a primary goal of bispectrum analyses, are studied in \cite{FlossBiagettiMeerburg2023}. Furthermore, these same squeezed configurations have the largest correlation with power spectrum measurements (see \cite{NovellMasotGilMarinVerde2023} for a recent assessment of the diagonal approximation and relevance of non-Gaussian contributions). 

A theoretical model, based on Perturbation Theory, of the covariance of redshift-space power spectrum and bispectrum is considered in \cite{GualdiEtal2018, GualdiEtal2019} as an ingredient of their compression method and in \cite{GualdiVerde2020} as the basis for a hybrid covariance where few parameters are calibrated against simulations. The model includes the non-Gaussian contributions expected for measurements in boxes with periodic boundary conditions but no additional source of covariance due to super-sample fluctuations \cite{HamiltonRimesScoccimarro2006, SefusattiEtal2006, DePutterEtal2012, TakadaHu2013}. The difference with the full numerical estimate, based on the PATCHY mocks of \cite{KitauraEtal2016} that does include survey geometry effects, is provided in terms of constraints on the model parameters showing negligible differences for a data vector including power spectrum monopole and quadrupole and bispectrum monopole. 

A similar analytical model is adopted in \cite{SugiyamaEtal2020}. In its evaluation all quadratic local and nonlocal bias terms are neglected, along with any modeling Finger-of-God effects. In addition, they also do not account for survey geometry effects, but still compare their predictions with the same PATCHY mocks that include observational effects. Despite the limitations of the model, the agreement at the level of the covariance matrix is qualitatively rather good, particularly for the statistics autocorrelations. This is because shot-noise represents the largest contribution where higher-order corrections are neglected. Indeed, the most severe discrepancies are observed for the cross-covariance between power spectrum and bispectrum and between the bispectrum monopole and quadrupole.  

In this work we focus our attention on the non-Gaussian contributions to the power spectrum and bispectrum cross-covariance and bispectrum covariance that are expected even in the absence of super-sample effects, extending to redshift space the results of \cite{BiagettiEtal2022} (see \cite{ChanMoradinezhadNorena2018} for a first take on super-sample bispectrum covariance). Unlike \cite{GualdiVerde2020} that follows the bispectrum estimator definition of \cite{HashimotoRaseraTaruya2017} and \cite{SugiyamaEtal2020} that adopt the decomposition introduced in \cite{SugiyamaEtal2019}, we consider the estimator choice of \cite{ScoccimarroCouchmanFrieman1999} and \cite{Scoccimarro2015}. We study the validity of the approximation of \cite{BiagettiEtal2022} for the non-Gaussian bispectrum covariance against a precise numerical estimate based on a very large set of mock halo catalogs. While this will not constitute a full description of the power spectrum and bispectrum errors in the setting of a realistic survey, we expect it to provide a sufficiently accurate estimate of one of its multiple contribution.

This paper is organized as follows: In section~\ref{sec:theory} we present a theoretical description of the full covariance of redshift-space power spectrum and bispectrum multipoles, including the approximation tested in \cite{BiagettiEtal2022} in real space. In section~\ref{sec:comparison} we provide a comparison of our prescription with a precise numerical estimate of the covariance 
and check the quality of the analytical inverse covariance matrix. Finally, we draw our conclusions in section~\ref{sec:conclusions}.

\section{Power spectrum and bispectrum covariance in redshift space}
\label{sec:theory}

In this section, we review the theoretical modeling of the covariance of power spectrum and bispectrum multipoles in redshift space along with their  cross-covariance. We will not consider any super-sample contribution to these quantities, expected in actual galaxy surveys due to finite-volume effects. While our models will correspond to contributions also present in real-survey observations, they only constitute a complete description of the power spectrum and bispectrum covariance properties as measured in simulation boxes with periodic boundary conditions. 

For this reason, we adopt the following definition for a generic $N$-point correlation function in Fourier space, $P_N$, accounting explicitly for the discrete nature of the wavenumbers $\kv$, 
\begin{align} \label{eq:N-point}
    \lan \delta(\kv_1)\cdots\delta(\kv_N)\ran \equiv \frac{\delta_K(\kv_{1\dots N})}{k_f^3} P_N(\kv_1,\dots,\kv_N)\,.
\end{align}
Here we adopt the notation $\kv_{1\dots N} = \kv_1 + \cdots + \kv_N$ and assume a finite cubic volume leading to the wavenumbers values $\kv=\nv\, k_f$,  multiples of the fundamental frequency $k_f\equiv 2\pi/L$, with $L$ being the side of the box. We also introduce the the adimensional Kronecker symbol $\delta_K(\kv)=1$ for $\kv=0$, vanishing otherwise. We work in the plane-parallel approximation throughout.

\subsection{Power spectrum estimator and covariance}

The covariance of the power spectrum has already been studied, in its Gaussian and non-Gaussian components, in several works both for measurements in simulation boxes with periodic boundary conditions \cite{ScoccimarroZaldarriagaHui1999, MeiksinWhite1999, TakahashiEtal2009, BertoliniEtal2016, GriebEtal2016, MohammedSeljakVlah2017, SugiyamaEtal2020, TaruyaMishimichiJeong2021} and in the more general case of a realistic galaxy survey including finite volume effects (see \citep{WadekarScoccimarro2020} and references therein). 

We assume the standard estimator for the power spectrum multipoles given by (see e.g. \cite{Scoccimarro2015})
\begin{equation}
\label{eq:PS estimator}
    \hat{P}_\ell(k) = (2\ell+1)\frac{k_f^3}{N_k} \sum_{\qv\in k} \delta_s(\qv)\,\delta_s(-\qv)\,\mathcal{L}_{\ell}(\mu_\qv)\,,
\end{equation}
where $\delta_s$ is the redshift-space galaxy (or halo) number density in Fourier space, $\mathcal{L}_{\ell}$ a Legendre polynomial with $\mu_\qv=\hat{q}\cdot\hat{n}$ being the cosine of the angle between $\qv$ and the line of sight $\hat{n}$, while 
\begin{equation}
\label{eq:Fundamental_modes}
    N_k =  \sum_{\qv\in k} \simeq \frac{1}{k_f^3}\int_k \de^3 q = \frac{4\pi k^2 \Delta k}{k_f^3} \,
\end{equation}
is the number of modes inside the $k$-bin of radial size $\Delta k$, defined as $k-\Delta k/2<|\qv|\le k+\Delta k/2$. The result in the above equation holds in the continuum limit approximation, valid as long as many grid modes are present in each shell.

The covariance associated with the power spectrum multiple estimator is defined as
\begin{equation}
\cov^P_{\ell_1\ell_2}(k_i,k_j)\equiv\langle[\hat{P}_{\ell_1}(k_i)-\langle\hat{P}_{\ell_1}(k_i)\rangle][\hat{P}_{\ell_2}(k_j)-\langle\hat{P}_{\ell_2}(k_j)\rangle]\rangle\,,
\end{equation}
and, as is well known, it comprises a Gaussian contribution $\cov^{P\,(PP)}$, dominant at large scales (see e.g. \cite{GriebEtal2016}) and only depending on the power spectrum, plus an additional non-Gaussian component $\cov^{P\,(T)}$ sourced by a non-vanishing trispectrum \cite{ScoccimarroZaldarriagaHui1999}.

An explicit expression for the Gaussian component is given by \cite{GriebEtal2016} 
\begin{equation}
    \cov^{P\,(PP)}_{\ell_1\ell_2}(k_i,k_j) = \frac{(2\ell_1+1)(2\ell_2+1)}{N_{k_i} N_{k_j}} \delta^K_{ij}\sum_{\qv\in k_i} P_{\tot}^2(\qv)\, \mathcal{L}_{\ell_1}(\mu_{\qv})\left[\mathcal{L}_{\ell_2}(\mu_{\qv})+\mathcal{L}_{\ell_2}(-\mu_{\qv})\right]\,,
\end{equation}
where $P_{\tot}(\kv)$ is the anisotropic halo power spectrum {\em including} the shot noise contribution equal, in the Poisson limit, to $1/[(2\pi)^3\bar{n}]$, with $\bar{n}$ the galaxy number density. Expanding the power spectrum in multipoles as $P_{\tot}(\qv)=\sum_{\ell}P_{\tot,\ell}(q)\mathcal{L}_{\ell}(\mu)$, this last equation can be written in the continuum limit as
\begin{align}
    \cov^{P\,(PP)}_{\ell_1\ell_2}(k_i,k_j) = & \,\frac{(2\ell_1+1)(2\ell_2+1)}{2N_{k_i}} \delta^K_{ij}\sum_{\ell_3=0}^{\infty}\sum_{\ell_4=0}^{\infty} P_{\tot,\ell_3}(k_i)P_{\tot,\ell_4}(k_i)\nonumber\\
    & \times \int^{1}_{-1}\,d\mu\,\mathcal{L}_{\ell_1}(\mu)\left[\mathcal{L}_{\ell_2}(\mu)+\mathcal{L}_{\ell_2}(-\mu)\right]\mathcal{L}_{\ell_3}(\mu)\mathcal{L}_{\ell_4}(\mu)\,.
\end{align}
where we used the thin-shell approximation assuming $P(q)\sim P(k)$. Since higher-order multipoles are suppressed, it is typically a good approximation to include only the first few terms in this sum.

In this work, we will ignore the non-Gaussian contribution to the power spectrum covariance as it is beyond its scope. Furthermore, while in some cases relevant, it is subdominant w.r.t. to the Gaussian component at large scales,  (see, e.g. \cite{WadekarScoccimarro2020}), unlike what happens in the case of the bispectrum.

\subsection{Bispectrum estimator and covariance}
\label{sec:theory-bisp}

A theoretical expression for the full bispectrum covariance excluding finite-volume effects is given in  \cite{SefusattiEtal2006} with specific expressions for the redshift-space multipoles presented in \cite{GualdiVerde2020, SugiyamaEtal2020, IvanovEtal2023, RizzoEtal2023}. In particular, \cite{RizzoEtal2023} provides a detailed comparison of its Gaussian approximation with numerical estimates, while a more complete test, including non-Gaussian contributions, has been considered in \cite{SugiyamaEtal2020}. 

\begin{figure}
\begin{center}
\includegraphics[scale=0.7]{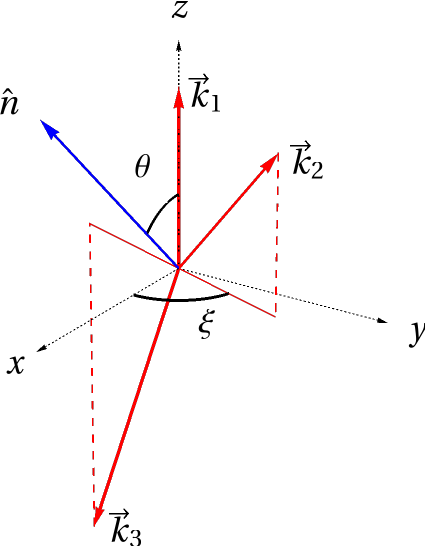}
\caption{Naming convention and axis orientation for the vectors and angles in the harmonic decomposition of equation~\eqref{eq:harmonic}.}
\label{fig:vectors}
\end{center}
\end{figure}

The expansion in spherical harmonics of the anisotropic bispectrum adopts the convention of \cite{ScoccimarroCouchmanFrieman1999, Scoccimarro2015}:
\begin{equation}
\label{eq:harmonic}
B_s(\kv_1, \kv_2, \kv_3, \hat{n}) \equiv \sum_{\ell=0}^{\infty}\sum_{m=-\ell}^\ell \,B_{\ell}^{m}(k_1, k_2, k_3)\,Y_{\ell}^m (\mu_{\kv_1}, \xi_{\kv_{12}})\,,
\end{equation}
where $\mu_{\kv_1} = \hat{k}_1\cdot\hat{n}$ and 
$\xi_{\kv_{12}}$ is the azimuthal angle of $\kv_2$ around $\kv_1$ with respect to the plane formed by $\kv_1$ and $\hat{n}$ as shown in Figure~\ref{fig:vectors}. Furthermore we define 
\begin{equation}
B_\ell (k_1, k_2, k_3)\equiv \sqrt{\frac{2\ell + 1}{4\pi}}B_{\ell}^{0}(k_1, k_2, k_3)\,,
\label{eq:harmonic_0}
\end{equation}
and we will refer to these quantities as bispectrum multipoles in the rest of the paper. We will ignore the case $m\ne 0$ thereby restricting ourselves to averages over the azimuthal angle $\xi$ (see \cite{GagraniSamushia2017} for an estimate of the additional information in higher-$m$ multipoles).  We will also assume, with no (further) loss of generality, the vector $\kv_1$ to be the largest of the triplet.\footnote{We do notice that the opposite choice, i.e. $\kv_1$ being the smallest, could lead to simpler expressions for our final results. However, this would lead, given the approximations that we will consider in its evaluation, to more severe systematic effects related to the discrete nature of the wavenumbers.}

In the plane-parallel approximation, the  bispectrum multipoles estimator is then given by
\begin{equation} \label{eq:Bs estimator}
    \hat{B}_\ell(t) =(2\ell+1) \frac{k_f^3}{N_t}\sum_{\qv_1\in k_1} \sum_{\qv_2 \in k_2} \sum_{\qv_3 \in k_3} \delta_K(\qv_{123}) \delta_s(\qv_1) \delta_s(\qv_2) \delta_s(\qv_3) \mathcal{L}_{\ell}(\mu_{\qv_1})\,,
\end{equation}
where $t$ represents the wavenumbers triplet $(k_1, k_2, k _3)$. We will refer to it as the ``triangle bin'', with the ``fundamental'' (i.e. located on a $k_f$-spaced grid) wavenumbers $\qv_i$ satisfying the relations $k_i-\Delta k/2\leq|\qv_i|<k_i+\Delta k/2$ for all $i=1$, 2 and 3 and ensuring the closed triangle condition as $\qv_{123}=0$. The normalization factor corresponds to the number of fundamental triplets $(\qv_1,\qv_2,\qv_3)$ falling in the triangle bin $(k_1, k_2, k _3)$, given by
\begin{equation}
\label{eq:fundamental triangles}
    N_t = \sum_{\qv_1\in k_1} \sum_{\qv_2 \in k_2} \sum_{\qv_3 \in k_3} \delta_K(\qv_{123}) \,.
\end{equation}

The bispectrum multipoles covariance is defined as 
\begin{equation}
\label{eq:covariance_def}
    \cov^B_{\ell_1\ell_2}(t_i,t_j) \equiv \langle[\hat{B}_{\ell_1}(t_i)-\langle\hat{B}_{\ell_1}(t_i)\rangle][\hat{B}_{\ell_2}(t_j)-\langle\hat{B}_{\ell_2}(t_j)\rangle]\rangle\,.
\end{equation} 
It can be split into a Gaussian and three, distinct non-Gaussian contributions \cite{SefusattiEtal2006, SugiyamaEtal2020}
\begin{equation}
\label{eq:covariance_full}
    \cov^B_{\ell_1\ell_2}(t_i,t_j)  = \cov_{\ell_1\ell_2}^{B\, (PPP)}(t_i,t_j) +\cov_{\ell_1\ell_2}^{B\,(BB)}(t_i,t_j) + \cov_{\ell_1\ell_2}^{B\,(PT)}(t_i,t_j) + \cov_{\ell_1\ell_2}^{B\,(P_6)}(t_i,t_j)\,,
\end{equation} 
where $\cov^{B\,(PPP)}$ refers to the Gaussian covariance, $\cov^{B\,(BB)}$ and $\cov^{B\,(PT)}$ refer to the disconnected non-Gaussian contributions containing two bispectra and a trispectrum respectively, and $\cov^{B\,(P_6)}$ refers to the connected six-point function contribution.

\subsubsection*{Gaussian contribution}
\label{sec:gaussian-covariance}

The Gaussian term in the bispectrum covariance can be written as \cite{RizzoEtal2023, IvanovEtal2023}
\begin{align}
    \cov_{\ell_1\ell_2}^{B\,(PPP)}(t_i,t_j)
    = & \frac{(2\ell_1+1)(2\ell_2+1)}{N_{t_i}N_{t_j}k_f^3}\delta^K_{ij} \sum_{\qv_1\in k_{1,i}} \sum_{\qv_2 \in k_{2,i}} \sum_{\qv_3 \in k_{3,i}} \delta_K(\qv_{123}) P_{\rm tot}(\qv_1)P_{\rm tot}(\qv_2)P_{\rm tot}(\qv_3)\nonumber\\
    & \times\left[\,(1+\delta^K_{k_{2,i},k_{3,i}})\mathcal{L}_{\ell_1}(\mu_{\qv_1})\mathcal{L}_{\ell_2}(-\mu_{\qv_1})\right.\nonumber\\
    &\left.\quad\;\; +(\delta^K_{k_{1,i},k_{2,i}}+\delta^K_{k_{1,i},k_{2,i}}\delta^K_{k_{2,i},k_{3,i}})\mathcal{L}_{\ell_1}(\mu_{\qv_1})\mathcal{L}_{\ell_2}(-\mu_{\qv_2})\right.\nonumber\\
    &\left.\quad\;\; +(\delta^K_{k_{1,i},k_{3,i}}+\delta^K_{k_{1,i},k_{2,i}}\delta^K_{k_{2,i},k_{3,i}})\mathcal{L}_{\ell_1}(\mu_{\qv_1})\mathcal{L}_{\ell_2}(-\mu_{\qv_3})\,\right]\,,
\end{align}
for $\ell_1=\ell_2 = 0$ the terms in the squared parenthesis reduce to a factor equal to $6$, 2 and 1 for equilateral, isosceles and scalene triangles, respectively. Expanding the anisotropic power spectrum in multipoles and using the thin-shell approximation, that is evaluating functions of the momenta at the bin centers, the Gaussian covariance becomes
\begin{align} 
\label{eq:C_PPP}
    \cov_{\ell_1\ell_2}^{B\,(PPP)}(t_i,t_j) 
    = & \; \frac{(2\ell_1+1)(2\ell_2+1)}{N_{t_i}k_f^3} \delta^K_{ij}\sum_{\ell_3,\ell_4,\ell_5} P_{{\rm tot},\ell_3}(k_{1,i})P_{{\rm tot},\ell_4}(k_{2,i})P_{{\rm tot},\ell_5}(k_{3,i})\nonumber\\
    & \times  R_{\ell_1,\ell_2,\ell_3,\ell_4,\ell_5}(k_{1,i},k_{2,i},k_{3,i})\,,
\end{align}
where
\begin{align}
    \label{eq:R}
    R_{\ell_1,\ell_2,\ell_3,\ell_4,\ell_5}(t) 
    =& \, \frac{1}{N_{t}} \sum_{\qv_1\in k_1} \sum_{\qv_2 \in k_2} \sum_{\qv_3 \in k_3} \delta_K(\qv_{123}) \mathcal{L}_{\ell_3}(\mu_{\qv_1})\mathcal{L}_{\ell_4}(\mu_{\qv_2})\mathcal{L}_{\ell_5}(\mu_{\qv_3})\nonumber\\
    & \times\left[\,(1+\delta^K_{k_{2,i},k_{3,i}})\mathcal{L}_{\ell_1}(\mu_{\qv_1})\mathcal{L}_{\ell_2}(-\mu_{\qv_1})\right.\nonumber\\
    &\left.\quad\;\;+(\delta^K_{k_{1,i},k_{2,i}}+\delta^K_{k_{1,i},k_{2,i}}\delta^K_{k_{2,i},k_{3,i}})\mathcal{L}_{\ell_1}(\mu_{\qv_1})\mathcal{L}_{\ell_2}(-\mu_{\qv_2})\right.\nonumber\\
    &\left.\quad\;\;+(\delta^K_{k_{1,i},k_{3,i}}+\delta^K_{k_{1,i},k_{2,i}}\delta^K_{k_{2,i},k_{3,i}})\mathcal{L}_{\ell_1}(\mu_{\qv_1})\mathcal{L}_{\ell_2}(-\mu_{\qv_3})\,\right] \,.
\end{align}
As long as there are many grid modes in each shell, this expression can be simplified in the continuum limit by replacing the sum with integrals, yielding an approximated analytical formula for the Gaussian covariance. The procedure and formula are detailed in equation~\eqref{eq:apx:fullPPP} and Appendix~\ref{sec:apx:continuum}. For this procedure to work the triangles need to be closed of flattened, i.e. to satisfy $k_1 > k_2 + k_3$ or $k_1 = k_2 + k_3$ respectively. Note that in the latter case, the continuum assumption may not be justified as flattened shells contain less modes than their closed counterpart, especially if one or more sides are small. As per the case of the power spectrum, we disregard the contribution of high-$\ell$ multipoles. In fact, we only consider $\ell_{3, 4, 5} = 0, 2$ and the results in \S\ref{subsec:check_cov} will show that no higher mixing is needed to accurately model the covariance of both the monopole and quadrupole.

\subsubsection*{Non-Gaussian contributions}
\label{sec:non-gaussian-covariance}

We begin by considering the non-Gaussian contribution depending on the product of two bispectra, that is 
\begin{multline}
    \cov_{\ell_1\ell_2}^{B\,(BB)}(t_i,t_j) =  \frac{(2\ell_1+1)(2\ell_2+1)}{N_{t_i}\,N_{t_j}} \sum_{\qv\mathrm{'s}\in t_i}\sum_{\pv\mathrm{'s}\in t_j} \delta_K(\qv_{123})\delta_K(\pv_{123}) \delta_K(\qv_{3}+\pv_{12})\delta_K(\pv_{3}+\qv_{12}) \\
    \times B_{\rm tot}(\qv_1,\qv_2,\pv_3)B_{\rm tot}(\pv_1,\pv_2,\qv_3) \mathcal{L}_{\ell_1}(\mu_{\qv_1})\mathcal{L}_{\ell_2}(\mu_{\pv_1})+ 8\ \text{perm.}\,,
\end{multline}
where we used the short-hand notation
\begin{equation}
\sum_{\qv\mathrm{'s}\in t}\equiv \sum_{\qv_1\in k_1}\sum_{\qv_2\in k_2}\sum_{\qv_3\in k_3}\,,
\end{equation}
for the sum over all values for the vectors $\qv_i$ in the triangle bin $t=(k_1,k_2,k_3)$. $B_\tot(\kv_1,\kv_2,\kv_3)$ is the anisotropic bispectrum including the shot noise contribution.
It should be noted that the first term on the r.h.s. is non-vanishing only if $k_{3,i}=k_{3,j}$ while the other eight permutations correspond to all other possible equalities between one element of the triplet $t_i$ and one of the triplet $t_j$. 

Using the harmonic decomposition of the anisotropic bispectra, as well as the thin-shell approximation we can write 
\begin{multline}
\label{eq:CBBfull}
\cov_{\ell_1\ell_2}^{B\,(BB)}(t_i,t_j) =  (2\ell_1+1)(2\ell_2+1) \sum_{\ell_3,m_3}\sum_{\ell_4,m_4} \,  B_{{\tot},\ell_3}^{m_3}(k_{1,i}, k_{2,i}, k_{3,j})\,B_{\tot,\ell_4}^{m_4}(k_{1,j}, k_{2,j}, k_{3,i}) \\ \times S^{(3,3)}_{\ell_1,\ell_2,\ell_3,\ell_4; m_3,m_4}(t_i,t_j) + 8\,\text{perm.}\,,
\end{multline}
where the mode-counting factor $S^{(a,b)}_{\ell_1,\ell_2,\ell_3,\ell_4; m_3,m_4}$ is defined as
\begin{multline}\label{eq:S}
     S^{(a, b)}_{\ell_1,\ell_2,\ell_3,\ell_4; m_3,m_4}(t_i,t_j) = \frac{1}{N_{t_i}N_{t_j}} \sum_{\qv\mathrm{'s}\in t_i}\sum_{\pv\mathrm{'s}\in t_j}\,\delta_K(\qv_{123})\,\delta_K(\pv_{123})\,\delta_K(\qv_a - \pv_b)  \\ 
      \times \mathcal{L}_{\ell_1}(\mu_{\qv_1})\mathcal{L}_{\ell_2}(\mu_{\pv_1}) Y^{m_3}_{\ell_3}(\mu_{\qv_1},\xi_{\qv_{12}})Y^{m_4}_{\ell_4}(\mu_{\pv_1},\xi_{\pv_{12}})\,.
\end{multline}

As pointed out in \cite{Barreira2019, BiagettiEtal2022}, the leading contribution to the non-Gaussian covariance comes from squeezed triangles that share the smallest momentum (long mode). In our prediction, we focus on this contribution and ignore all other permutations in equation~\eqref{eq:CBBfull}. We ignore the contribution coming from $m \neq 0$ multipoles as it is suppressed both by the amplitude of the multipoles of the bispectrum and by the angular integrals. The final expression for the covariance is
\begin{equation}\label{eq:finalBB}
    \cov^{B\,(BB)}_{\ell_1\ell_2}(t_i, t_j) = (2\ell_1 + 1) (2\ell_2 + 2) \delta^{K}_{k_{3, i}, k_{3, j}} \Sigma^{33}_{ij} \, \sum_{\ell_3, \ell_4} B_{\tot, \ell_3}(t_i)B_{\tot, \ell_4}(t_j) \, I_{\ell_1 \ell_2 \ell_3 \ell_4}(t_i, t_j)\,,
\end{equation}
where 
\begin{equation}
    I_{\ell_1 \ell_2 \ell_3 \ell_4}(t_i, t_j) = 4\pi \frac{S^{(3, 3)}_{\ell_1,\ell_2,\ell_3,\ell_4;0,0}(t_i, t_j)}{\Sigma_{ij}^{33}\sqrt{(2\ell_3 + 1)(2\ell_4 + 1)}}
\end{equation}
and the quantity $\Sigma_{ij}^{33} \equiv S^{(3, 3)}_{0,0,0,0;0,0}(t_i, t_j)$ was first defined in \cite{BiagettiEtal2022}. For closed triangles, it is equal in the continuum limit to 
\begin{equation}
    \Sigma^{33}_{ij} = \frac{16\pi^3}{k_f^9 N_{t_i} N_{t_j}} k_{1,i} k_{2,i} k_{1,j} k_{2,j} \Delta k^5\,,
\end{equation}
and expressions for non-closed triangles can be found in Appendix~A of \cite{BiagettiEtal2022}. The expression for $I_{\ell_1\ell_2\ell_3\ell_4}$ in the continuum limit is presented in equation~\eqref{eq:apx:fullBB} of Appendix \ref{sec:apx:continuum}. In the evaluation of equation~\eqref{eq:finalBB} for $\ell_{1, 2} = 0, 2$ we only consider contributions up to the quadrupole, that is we keep only $\ell_{3, 4} = 0, 2$.

Regarding the $\cov^{PT}$ contribution to the covariance from equation~\eqref{eq:covariance_full}, it is
\begin{multline}
    \cov_{\ell_1\ell_2}^{B\,(PT)}(t_i,t_j) =  \frac{(2\ell_1+1)(2\ell_2+1)}{N_{t_i}N_{t_j}} \sum_{\qv's\in\kv^is}\sum_{\pv's\in\kv^js} \delta_K(\qv_{123})\delta_K(\pv_{123}) \delta_K(\qv_{3}+\pv_{12})\delta_K(\pv_{3}+\qv_{12}) \\ 
    P_{\tot}(\qv_3)T_{\tot}(\qv_1,\qv_2,\pv_1,\pv_2) \mathcal{L}_{\ell_1}(\hat{q}_1\cdot\hat{n})\mathcal{L}_{\ell_2}(\hat{p}_1\cdot\hat{n})+ 8\ \text{perm.}\,.
\end{multline}
Again, we keep only the leading term, which is the permutation with the power spectrum evaluated on the long mode (i.e. the one written explicitly in this expression). For that permutation, 
it is argued in \cite{BiagettiEtal2022} that the products $P_{\tot}(\qv_3)\,T_{\tot}(\qv_1,\qv_2,\pv_1,\pv_2)$ and $B_{\tot}(\qv_1, \qv_2, \qv_3)\,B_{\tot}(\pv_1, \pv_2, \qv_3)$ have the same limit for squeezed triangles (taking $q_3 = |\qv_1 + \qv_2|$ much smaller than all other momenta).
The same argument holds in redshift space (see for example~\cite{CreminelliEtal2013, KehagiasEtal2013}). Since this contribution is expected to be important only for squeezed triangles, we account for it by writing $\cov^{B\,(PT)} \simeq \cov^{B\,(BB)}$.  We disregard the contribution coming from the connected six-point function, as argued in \cite{BiagettiEtal2022}. Thus, our non-Gaussian covariance is approximated by
\begin{equation}
\label{eq:cBB_NGmodel}
\cov^B_{\ell_1,\ell_2} \simeq \cov_{\ell_1\ell_2}^{B\,(PPP)} + 2\cov_{\ell_1\ell_2}^{B\,(BB)}
\end{equation}
where $\cov_{\ell_1\ell_2}^{B\,(BB)}$ is given by equation~\eqref{eq:finalBB}.

\subsection{Power spectrum-bispectrum cross-covariance}

Moving to the cross-covariance among power spectrum and bispectrum, this is defined as
\begin{equation}
\label{eq:covariance_cross}
    \cov^{PB}_{\ell_1,\ell_2}(k_i,t_j) \equiv \langle[\hat{P}_{\ell_1}(k_i)-\langle\hat{P}_{\ell_1}(k_i)\rangle][\hat{B}_{\ell_2}(t_j)-\langle\hat{B}_{\ell_2}(t_j)\rangle]\rangle\,,
\end{equation} 
and it comprises two terms
\begin{equation}
    \cov^{PB}_{\ell_1 \ell_2}(k_i, t_j) = \cov^{PB\,(PB)}_{\ell_1 \ell_2}(k_i, t_j) + \cov^{PB\,(P_5)}_{\ell_1 \ell_2}(k_i, t_j)\,,
\end{equation}
where $\cov^{PB\,(PB)}$ refers to the disconnected contribution, and $\cov^{PB\,(P_5)}$ refers to the fully-connected 5-point function contribution. Ignoring the latter, as it is subleading at large scales (see \cite{BiagettiEtal2022} and our numerical results later), the relevant term is
\begin{multline}
    \cov^{PB\,(PB)}_{\ell_1,\ell_2}(k_i,t_j) = \frac{(2\ell_1+1)(2\ell_2+1)}{N_{k_i}N_{t_j}} \, 2 \sum_{\qv \in k_i}\sum_{\qv's\in t_j} \delta_K(\qv_{123}) P_{\tot}(\qv) B_{\tot}(\qv_1,\qv_2,\qv_3) \\ 
    \times \left[\delta_K(\qv+\qv_1) + \delta_K(\qv+\qv_2)
    +\delta_K(\qv+\qv_3)\right] \mathcal{L}_{\ell_1}(\mu_{\qv}) \mathcal{L}_{\ell_2}(\mu_{\qv_1})\,.
\end{multline}
Similarly to what was done earlier, we decompose $P_{\rm tot}$ in Legendre polynomials and $B_{\rm tot}$ in spherical harmonics, disregarding $m \neq 0$. We apply the thin-shell approximation to get:
\begin{equation}\label{eq:fullPB}
    \cov^{PB\,(PB)}_{\ell_1,\ell_2}(k_i,t_j) = (2\ell_1+1)(2\ell_2+1)
    \sum_{\ell_3}\sum_{\ell_4} P_{\tot,\ell_3}(k_i) B_{\tot,\ell_4}(t_j)\,\Tilde{I}_{\ell_1\ell_2\ell_3\ell_4}(k_i, t_j)\,,
\end{equation}
where the quantity $\Tilde{I}_{\ell_1\ell_2\ell_3\ell_4}$ is defined as
\begin{multline}\label{eq:tildeI}
    \Tilde{I}_{\ell_1\ell_2\ell_3\ell_4}(k_i, t_j) = \frac{2}{N_{k_i}N_{t_j}}\sum_{\qv\mathrm{'s}\in t_j} \delta_K(\qv_{123})   \mathcal{L}_{\ell_2}(\mu_{\qv_1})\mathcal{L}_{\ell_4}(\mu_{\qv_1}) \\ 
    \times\left[\delta^K_{k_i,k_{1,j}} \mathcal{L}_{\ell_1}(\mu_{\qv_1})\mathcal{L}_{\ell_3}(\mu_{\qv_1}) + \delta^K_{k_i,k_{2,j}} \mathcal{L}_{\ell_1}(\mu_{\qv_2})\mathcal{L}_{\ell_3}(\mu_{\qv_2}) +\delta^K_{k_i,k_{3,j}} \mathcal{L}_{\ell_1}(\mu_{\qv_3})\mathcal{L}_{\ell_3}(\mu_{\qv_3}) \right]\,.
\end{multline}
Its expression in the continuum limit can be found in Appendix~\ref{sec:apx:continuum}, see equation~\eqref{eq:apx:fullPB}.

To sum up, our approximate model for the cross-covariance of power spectrum and bispectrum is 
\begin{equation}
\label{eq:cPB_NGmodel}
\cov^{PB}_{\ell_1,\ell_2} \simeq \cov_{\ell_1\ell_2}^{PB\,(PB)} \,
\end{equation}
with $\cov_{\ell_1\ell_2}^{PB\,(PB)}$ is given by the expression above.

\section{Validation: comparison with numerical results}
\label{sec:comparison}

\subsection{Data}

Our benchmark is a numerical estimate of the power spectrum and bispectrum covariance from $10000$ mock halo catalogs created with the \texttt{Pinocchio} code \cite{MonacoTheunsTaffoni2002, MonacoEtal2013, MunariEtal2017} implementing an approximate method to describe the halo distribution based on Lagrangian Perturbation Theory. These mocks share the same cosmology, box size, and resolution as the Minerva simulations of \cite{GriebEtal2016}, tracking the evolution of 1 billion dark matter particles in a box of size $L=1500\ h^{-1} \text{Mpc}$ per side. The reference Minerva halo catalogs are defined in terms of a minimum mass of $1.12 \times 10^{13}\ h^{-1} M_{\odot}$. The mass threshold in the \texttt{Pinocchio} mocks is set to closely match the amplitude of the large-scale halo power spectrum of the N-body simulations, including shot-noise. 

The \texttt{Pinocchio} halo catalogs were introduced for the real-space bispectrum analyses carried out in \cite{OddoEtal2020, OddoEtal2021} while the redshift-space measurements were made for  \cite{RizzoEtal2023} following equation~\eqref{eq:Bs estimator} for the bispectrum estimator. We consider all triangular configurations up to a $k_{\rm max} \simeq 0.12\kMpc$, thereby limiting ourselves to the regime where a tree-level perturbative description of the halo bispectrum is expected to be valid. As shown in \cite{RizzoEtal2023}, the bispectrum of the mocks is systematically underestimated (a few percent on these scales) with respect to the N-body simulations, as it is typical for approximate schemes based on Lagrangian perturbation theory. However, its variance is accurately reproduced since the dominant contribution to the covariance depends on the power spectrum, which is matched to the N-body results. Of course non-Gaussian contributions to the covariance that depend on the bispectrum or trispectrum are likely to be underestimated as well, but since the suppression amounts at most to a $\sim 10\%$ correction on these scales (and less for squeezed configurations), this corresponds  a sub-percent effect on the final covariance.

We will compare the numerical estimates of the bispectrum and cross covariance with the approximate model of equations \eqref{eq:cBB_NGmodel} and \eqref{eq:cPB_NGmodel}. Both expressions only require the evaluation of power spectrum and bispectrum multipoles, the latter even at tree-level in perturbation theory, because of the large-scales we are focusing on, and the large shot-noise contributions for our halo catalogs at small scales. However, as we are mainly interested in testing the validity of these models, we will make use of the average values $\langle\hat{P}_\ell(k)\rangle$ and $\langle\hat{B}_\ell(t)\rangle$ of the measurements obtained from all realizations to build our covariance matrix. This would account also for any effect due to the \texttt{Pinocchio} approximate displacements based on Lagrangian perturbation theory.

\subsection{Bispectrum covariance}
\label{subsec:check_cov}

\begin{figure}
    \centering
    \includegraphics[width=0.85\textwidth]{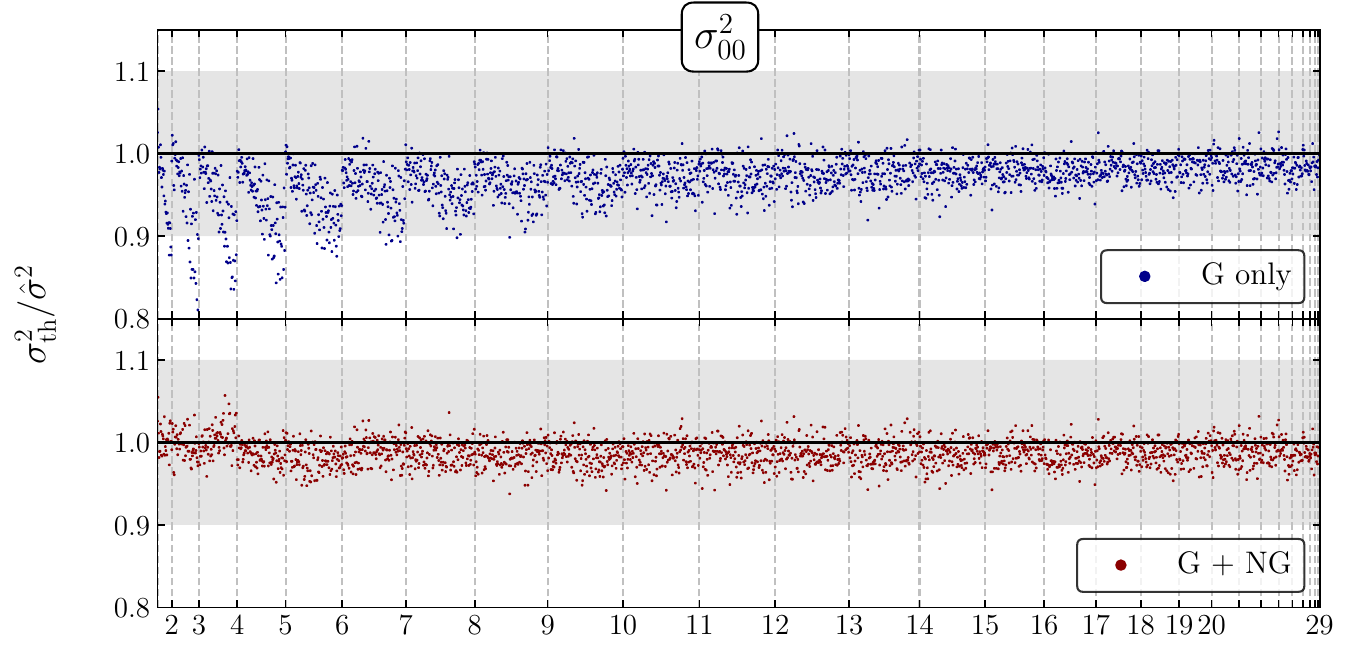}
    \includegraphics[width=0.85\textwidth]{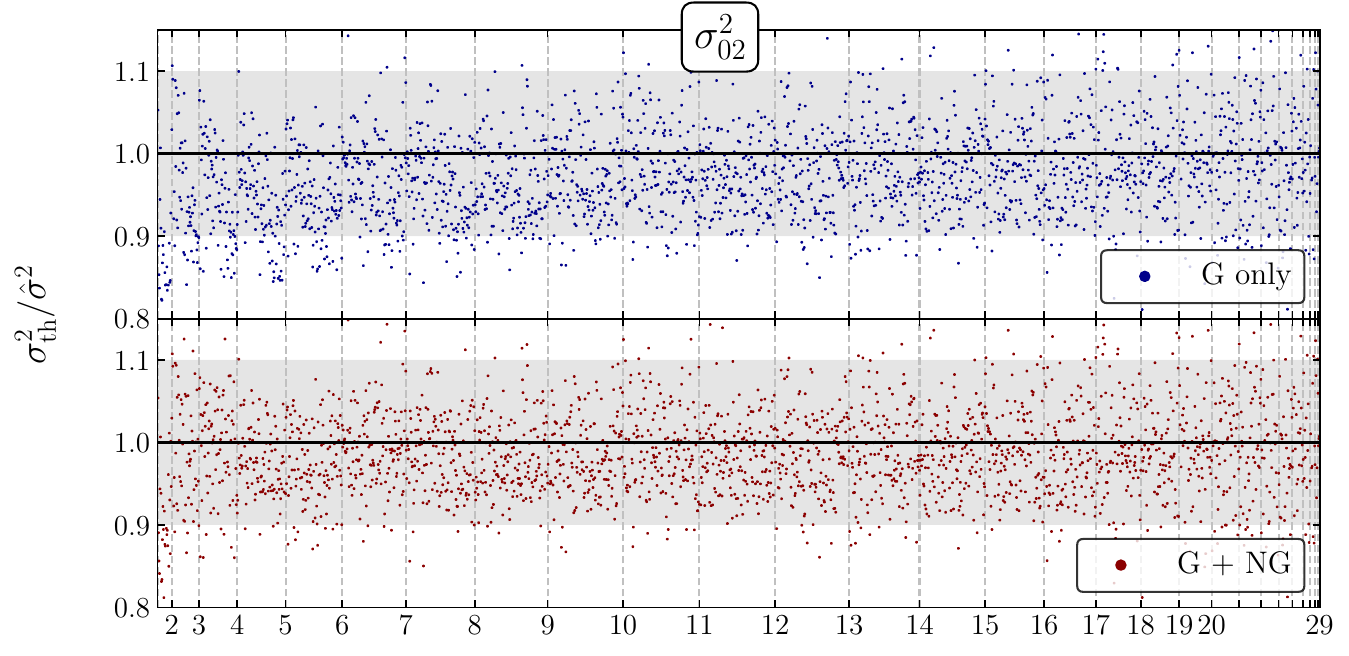}
    \includegraphics[width=0.85\textwidth]{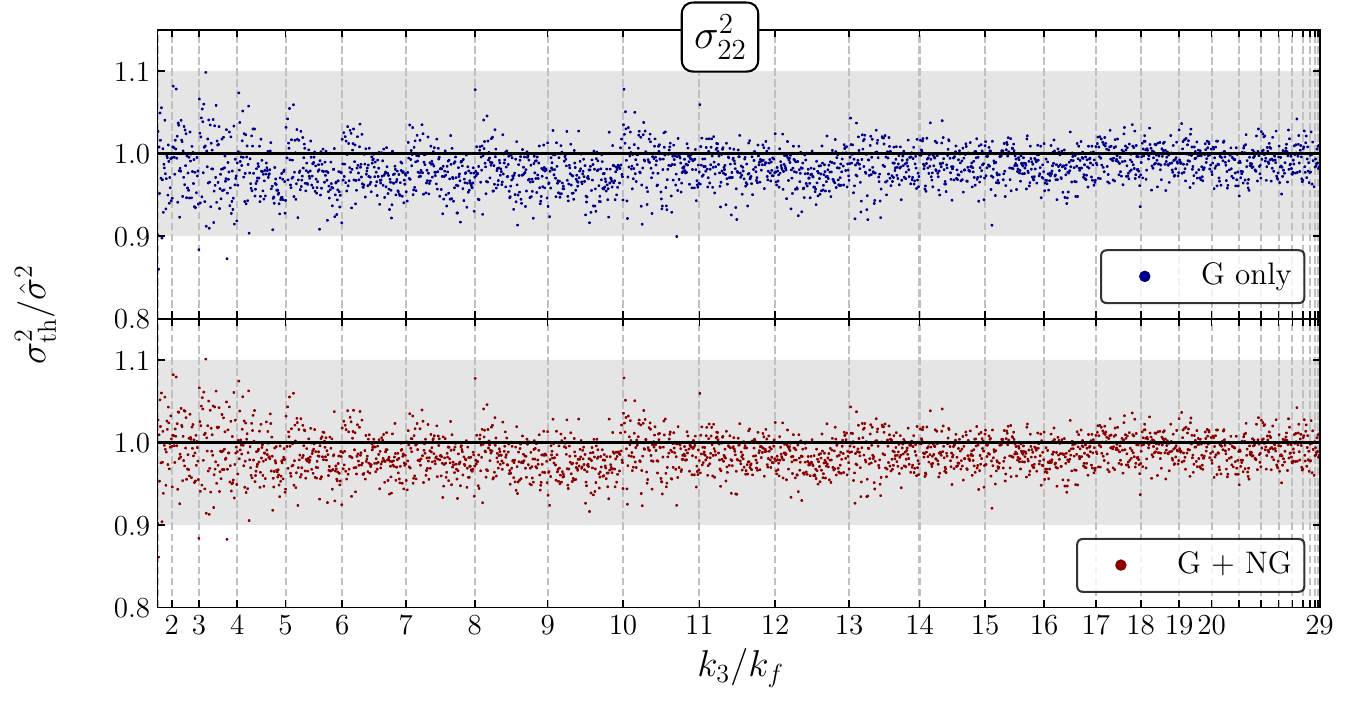}
    \caption{Bispectrum monopole variance $\sigma^2_{00}(t)$ (top panel), covariance between the different multipoles for the same triangular configuration $\sigma^2_{02}(t)$ (middle) and quadrupole variance $\sigma^2_{22}(t)$ (bottom). Each panel shows the ratio of the Gaussian prediction to the numerical estimate in the upper half, while in the lower half the model includes the non-Gaussian contributions according to equation \eqref{eq:cBB_NGmodel} as well. The triangles are ordered in groups sharing the same $k_3$ (smallest mode), then $k_2$, and finally $k_1$, for all values from $k_f \simeq 0.004\,h\,\text{Mpc}^{-1}$ up to $k_{\mathrm{max}} = 29k_f \simeq 0.12\,h\,\text{Mpc}^{-1}$.}
    \label{fig:bispectrum_variance}
\end{figure}

The three plots in Figure~\ref{fig:bispectrum_variance} show the ratio between the analytical and numerical covariance of bispectrum multipoles for the elements where we expect a Gaussian contribution. These correspond to the variance
\be
\sigma^2_{\ell\ell}(t)\equiv C^B_{\ell \ell}(t,t)\,,
\ee
along with the covariance between different multipoles for the same triangular configuration, that is $B_0(t)$ and $B_2(t)$, 
\be
\sigma^2_{02}(t)\equiv C^B_{02}(t,t)\,.
\ee
In each panel we show the comparison of both the Gaussian model and of the model including non-Gaussian contributions according to equation \eqref{eq:cBB_NGmodel}. Triangular configurations are ordered by increasing value of the smallest wavenumber $k_3$; then for each $k_{3}$ bin, delimited by vertical dashed lines in the figure, triangles are ordered by increasing $k_2$, and then $k_1$ (remember that the triangle sides satisfy $k_1 > k_2 > k_3$.) 

Let us focus on the monopole variance, $\sigma_{00}^2$. Very squeezed triangles are located at the left of the figure where most configurations have a small value for $k_{3}$, and toward the right for each bin, where the other two sides are larger. We see that the variance of such triangles is underestimated by the Gaussian approximation by more than $\gtrsim 10 \%$ in some cases, while the agreement is greatly improved when including the non-Gaussian term. It's worth noting that the model still systematically underestimates the true value of the covariance by around $2 \%$, as it is evident from the very last bins in the plot. This is the cumulative effect of all subleading non-Gaussian terms that are ignored in our approximation.
While the magnitude of non-Gaussian corrections is comparable across monopole and quadrupole, the Gaussian covariance is much larger for the latter, up to a factor $\gtrsim 5$ for squeezed configurations. Thus, non-Gaussian contributions appear to be relatively less important for higher multipoles.

We notice that the cross-covariance $\sigma_{02}^2$ displays the most noise. This is because its Gaussian part is dominated by the mixing between the power spectrum monopole and quadrupole $P_0 P_0 P_2$, equation \eqref{eq:C_PPP}, while in the other two cases, corresponding to the proper variance, the largest contribution comes from the $P_0 P_0 P_0$ term. 
These results agree with what was shown in~\cite{RizzoEtal2023}.\footnote{Note that we have a different ordering of the triangles. We also take the continuum approximation by replacing sums by integrals, while~\cite{RizzoEtal2023} perform the sums over the grid points inside each triangle bin. This gives a noticeable difference only for the configurations involving the smallest modes.}

\begin{figure}
    \centering
    \includegraphics[width=\textwidth]{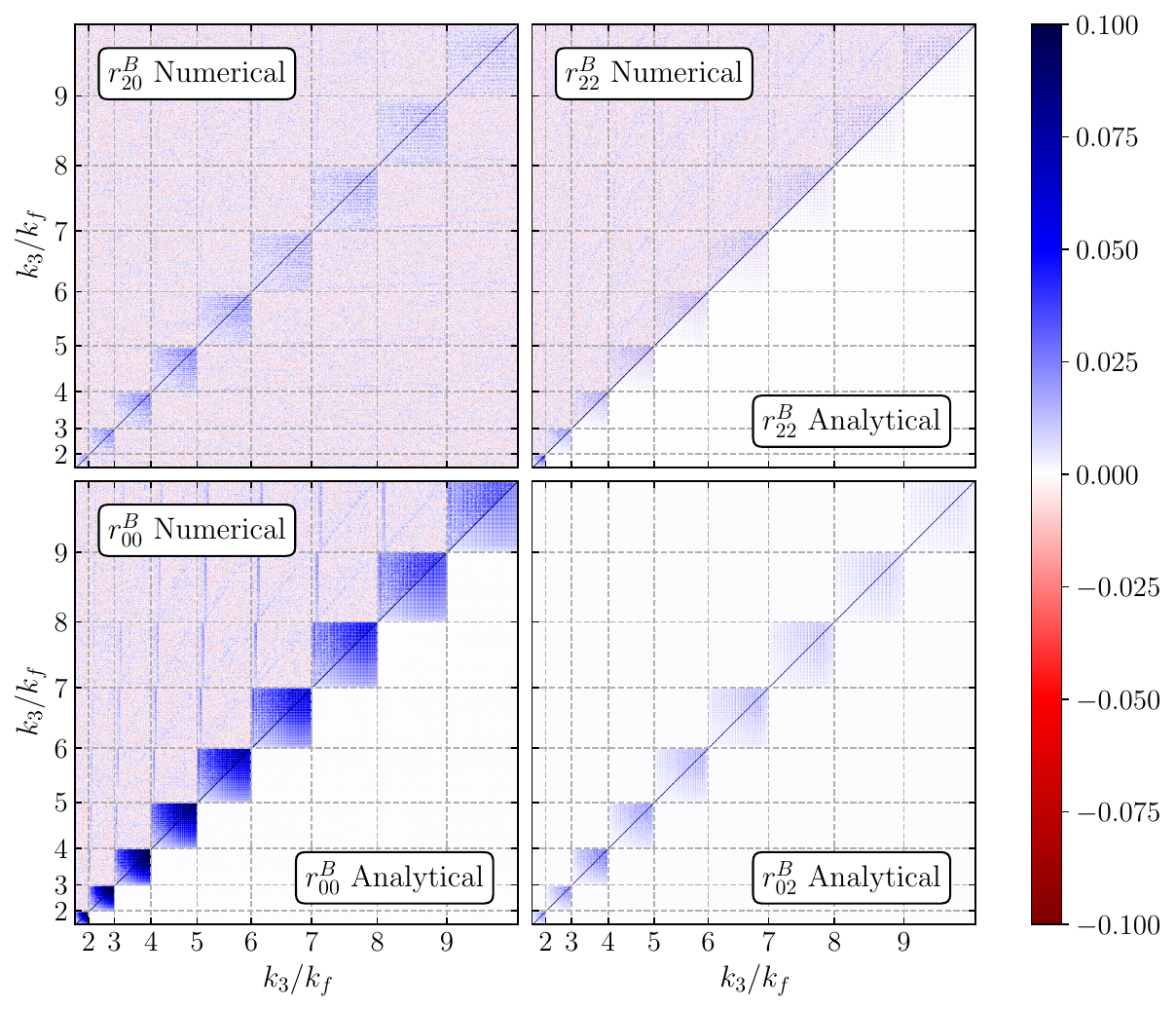}
    \caption{Bispectrum monopole and quadrupole correlation matrix, equation \eqref{eq:rBB}. 
    The triangles are ordered in groups that share the same $k_3$ (smallest mode), then $k_2$, and finally $k_1$. The axes labels mark increasing values of $k_3$, with all triangles with fixed $k_3$ appearing between each value.
    The upper-left part of the plot shows the estimate coming from the 10000 mocks while the model is plotted on the lower-right half.
    Modes range from $k_f \simeq 0.004\,h\,\text{Mpc}^{-1}$ up to $k_{\mathrm{max}} = 9k_f \simeq 0.04\,h\,\text{Mpc}^{-1}$.}
    \label{fig:BBcovariance}
\end{figure}

Figure~\ref{fig:BBcovariance} shows the same comparison for the whole correlation matrix, defined as
\begin{equation}
\label{eq:rBB}
    r^B_{\ell_1\ell_2}(t_i,t_j)\equiv\frac{\cov^B_{\ell_1\ell_2}(t_i,t_j)}{\sqrt{\cov^B_{\ell_1\ell_1}(t_i,t_i)\cov^B_{\ell_2\ell_2}(t_j,t_j)}}\,.
\end{equation}
Symmetric sub-matrices, i.e. $r_{00}$ and $r_{22}$, are split down the diagonal into its numerical and analytical estimates (top-left and bottom-right, respectively). For $r_{02}$ we show instead two separate plots: the top-left presenting the numerical estimate while the bottom-right the model prediction. 

The Gaussian model would be perfectly diagonal, and would miss the off-diagonal structure observed here, that is the blocks appearing along the diagonal. One can notice how the non-Gaussian terms of equation of \eqref{eq:cBB_NGmodel}  capture most of such structure. The largest covariance elements are between different squeezed triangle configurations sharing the same long mode. This is similar to what happens in real space \cite{BiagettiEtal2022}. As already noticed discussing the variance and Gaussian terms, we can see here as well the off-diagonal terms induced by non-Gaussianity are less important for higher multipoles, along with the fact that the covariance between different multipoles is noisier. We finally notice that some additional structure away from the main block-diagonal elements is not captured by our model. We will get back to this in section~\ref{sec:check}.

\subsection{Power spectrum bispectrum cross-covariance}

\begin{figure}
    \centering
    \includegraphics[width=0.9\textwidth]{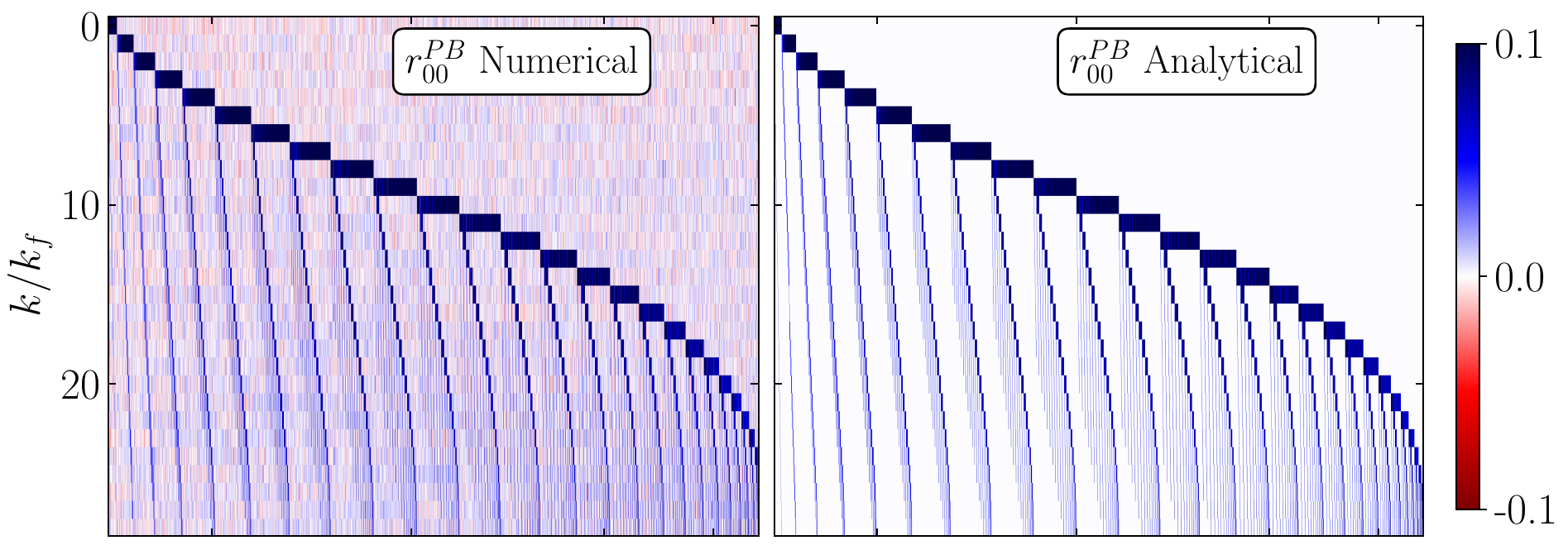}
    \includegraphics[width=0.9\textwidth]{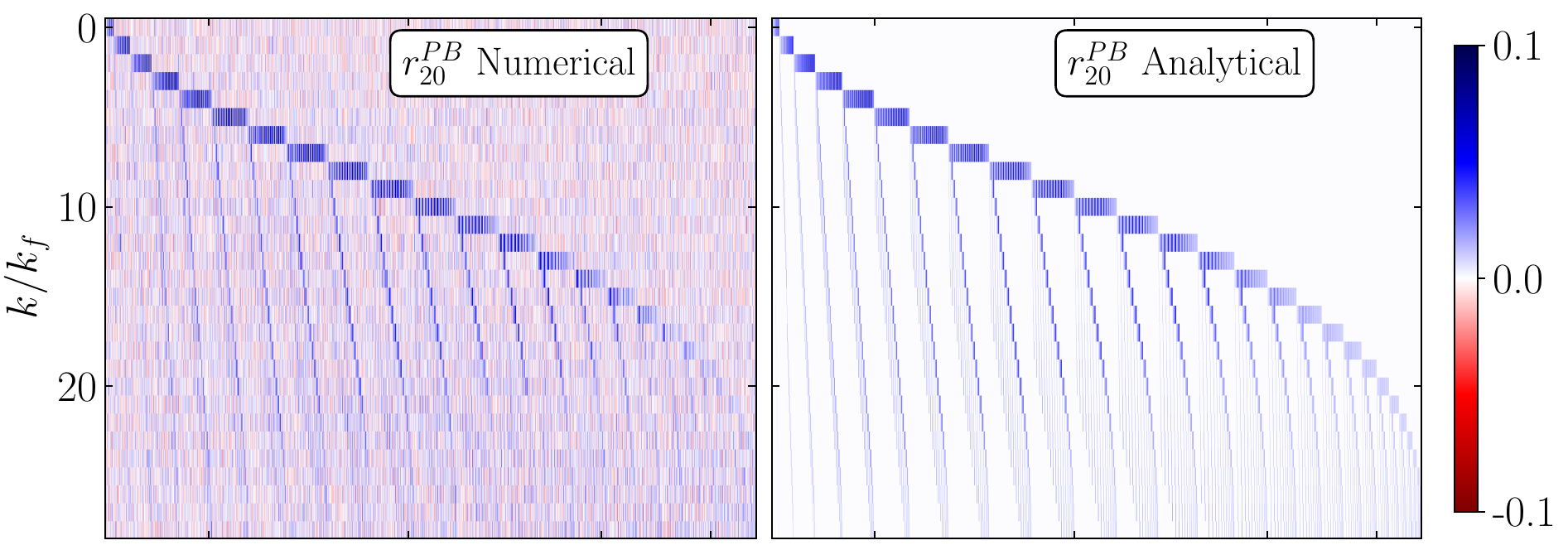}
    \includegraphics[width=0.9\textwidth]{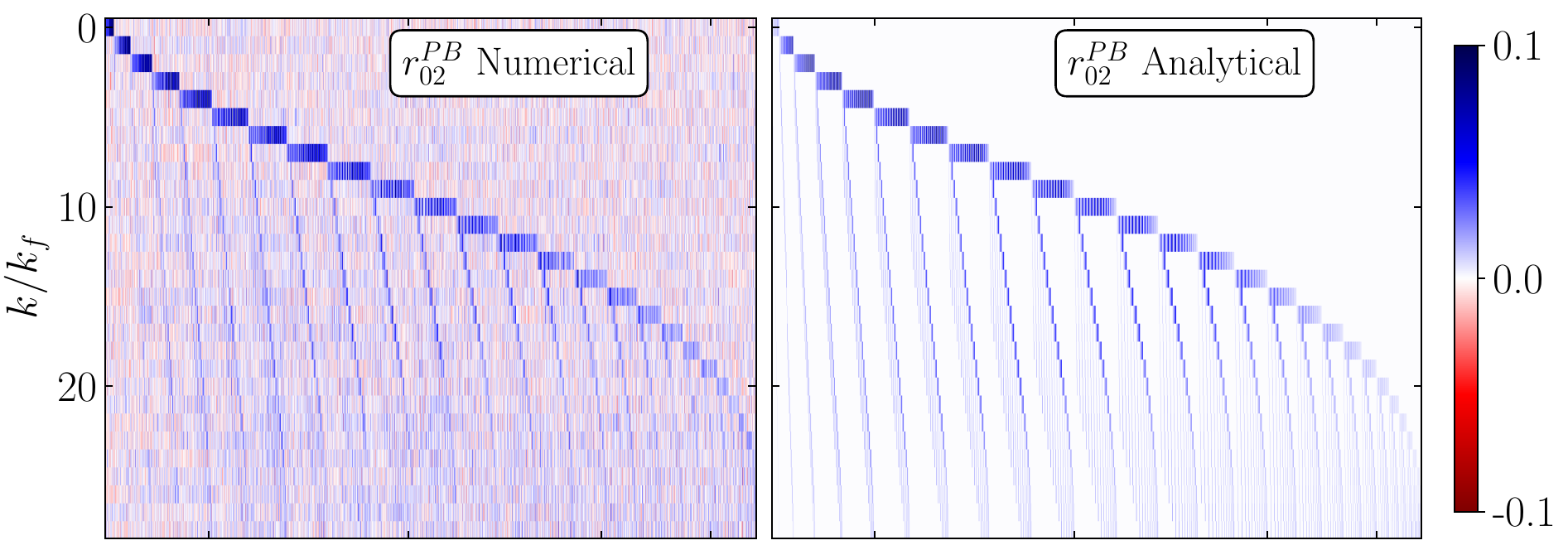}
    \includegraphics[width=0.9\textwidth]{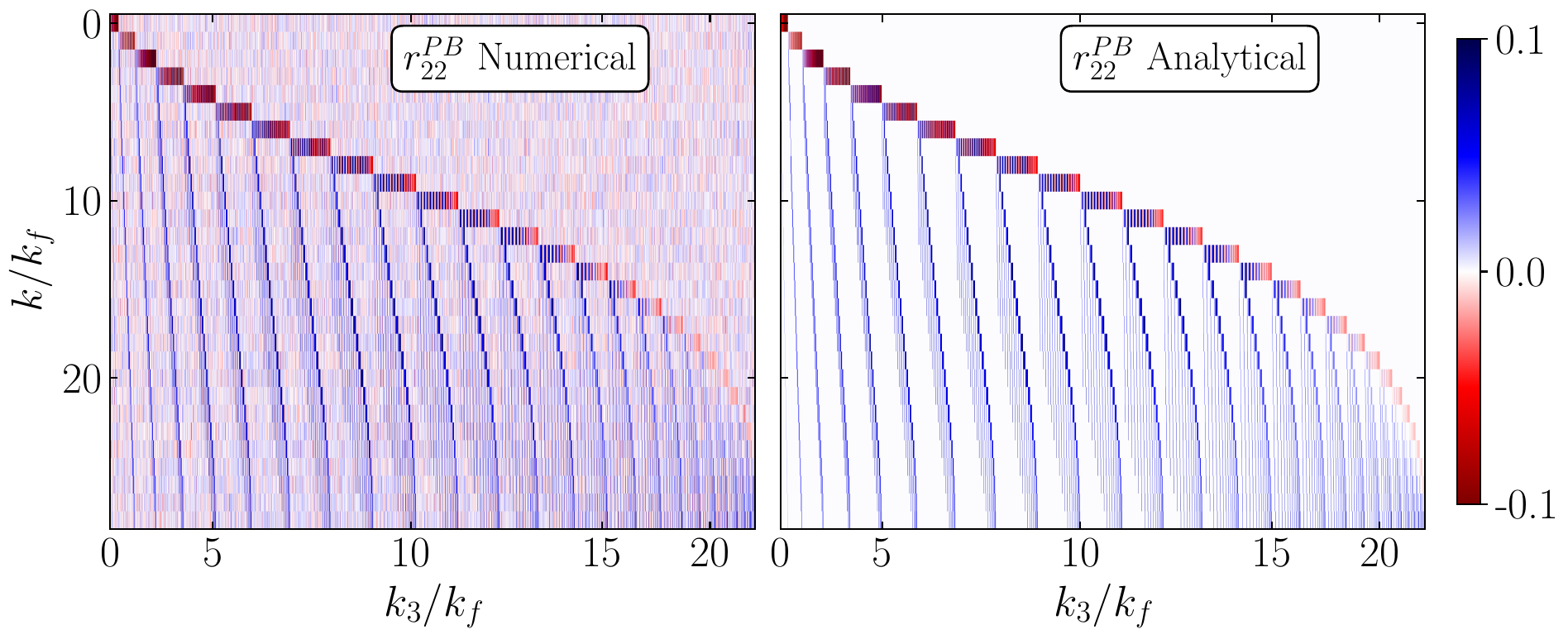}
    \caption{comparison of the power spectrum-bispectum correlation coefficients, equation \eqref{eq:rPB}, from the numerical estimate (left) to the model (right). Abscissae denote power spectrum modes while ordinates denote triangular bispectrum configurations. Modes range from $k_f \simeq 0.004\,h\,\text{Mpc}^{-1}$ up to $k_{\mathrm{max}} = 29k_f \simeq 0.12\,h\,\text{Mpc}^{-1}$.}
 \label{fig:PB cross correlation}
\end{figure}

We show in Figure~\ref{fig:PB cross correlation} the power spectrum-bispectrum cross-correlation coefficients, defined as
\begin{equation}
\label{eq:rPB}
    r^{PB}_{\ell_1\ell_2}(k_i,t_j)\equiv\frac{\cov^{PB}_{\ell_1\ell_2}(k_i,t_j)}{\sqrt{\cov^P_{\ell_1\ell_1}(k_i,k_i)\cov^B_{\ell_2\ell_2}(t_j,t_j)}}\,.
\end{equation}
As for the bispectrum covariance we observe that the main non-Gaussian features are accurately reproduced by the model, and, again, the monopole-quadrupole correlation is noisier while higher multipoles show less correlation. We see that the largest elements correspond to triangular configurations where the smallest side coincides with the power spectrum mode $k$, as already noted in real space~\cite{BiagettiEtal2022}. The model also reproduces the subdominant features due to  bispectrum triangles sharing the larger sides, $k_1$ and $k_2$, with the power spectrum.

\subsection{Tests on the inverse covariance}
\label{sec:check}

Approximate theoretical models for the covariance of complex data vectors such as the combination of power spectrum and bispectrum multipoles are not guaranteed to provide invertible covariance matrices. Indeed, imposing such (necessary) condition in general terms is not a simple problem. In practice we could expect that some level of consistency among the different components can help obtaining a well defined matrix. 

In this respect, we notice that using the expression for $\cov^{PB}(k_i, t_j)$ in equation~\eqref{eq:fullPB} and \eqref{eq:tildeI} returns a matrix with negative eigenvalues. This is likely related to the fact that these formulae consider all possible permutations when sharing the mode of the power spectrum with the bispectrum triangle, i.e. they account for $k_i = k_{1, j}, k_{2, j}, k_{3, j}$, whereas in the covariance of the bispectrum, $\cov^{BB}(t_i, t_j)$, we only account for the leading contribution corresponding to the triangles  sharing the long mode. We find that discarding the keeping only the term with  $k_i=k_{3, j}$ and discard the other two in the cross-covariance lead to an invertible matrix. Thus, for the results in this section we employ this simplified version of the cross-covariance model.

In order to assess the quality of our analytical covariance matrix and its inverse, we perform two tests, following \cite{BiagettiEtal2022}. For the first, we compute, for each realization $i$, the quantity
\begin{equation}
    \chi^2_{\mathrm{th}, i} = (\dv_i - \bar{\dv})\,\cov_{\mathrm{th}}^{-1}\,(\dv_i - \bar{\dv})^T\,,
\label{eq:chi2}
\end{equation}
where $\dv_i$ is the data vector containing the measured power spectrum and bispectrum monopole and quadrupole. If the model is a good description of the covariance, this quantity should follow a $\chi^2$ distribution with the number of degrees of freedom equal to the dimension of $\dv_i$.

\begin{figure}
    \centering
    \includegraphics[width=0.85\textwidth]{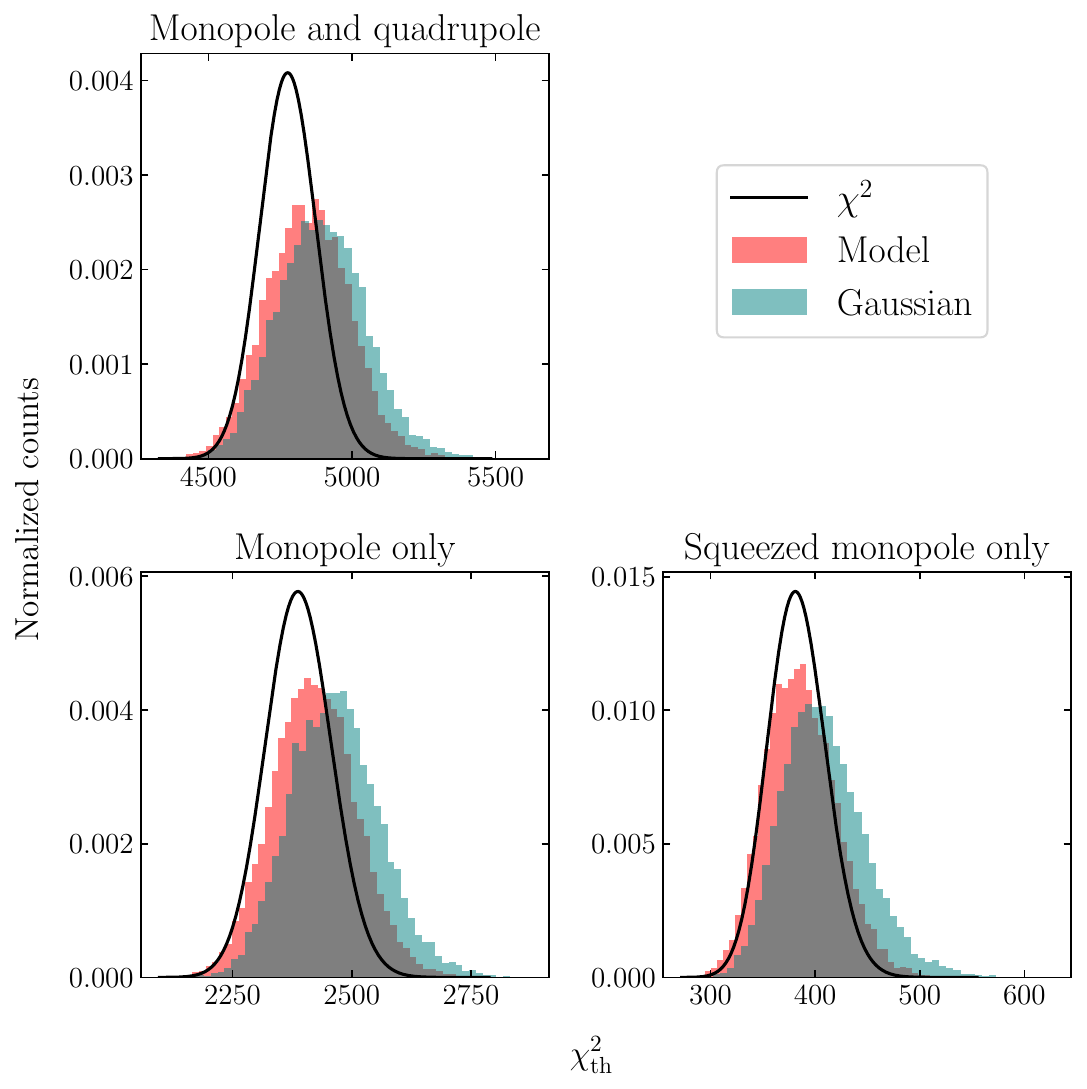}
    \caption{Distribution of $\chi^2_{\mathrm{th},i}$ values, as defined in equation~\eqref{eq:chi2}. The different results assuming Gaussian and non-Gaussian models for the covariance are shown in different colors while the expected $\chi^2$ distribution is shown with a solid line. The top panel shows the results considering all triangles for both monopole and quadrupole while the bottom panels correspond to a data vector limited to the monopole only (left) and to squeezed monopole configurations (right), the latter defined as triangles with $k_2 > 3\,k_3$.}
    \label{fig:chi2-test}
\end{figure}

In figure~\ref{fig:chi2-test}, we plot the histograms obtained for the distributions of this quantity using both the Gaussian covariance and the non-Gaussian model from section~\ref{sec:non-gaussian-covariance}, comparing it with the $\chi^2$ distribution. We consider the results for the full data vector along the cases corresponding to limiting the data vector to the bispectrum monopole and to the squeezed configurations of the monopole. We see that the non-Gaussian model consistently improves upon the Gaussian approximation. The effect is more pronounced for the monopole alone since, as discussed above, non-Gaussian contributions are less relevant for the quadrupole covariance. Furthermore, the last panel confirms that our model for non-Gaussian terms grows more accurate when looking at squeezed configurations. 
Notice also that the improvement is less dramatic than in \cite{BiagettiEtal2022} because we are considering less squeezed triangles in our mock measurements.

\begin{figure}
    \centering
    \includegraphics[width=\textwidth]{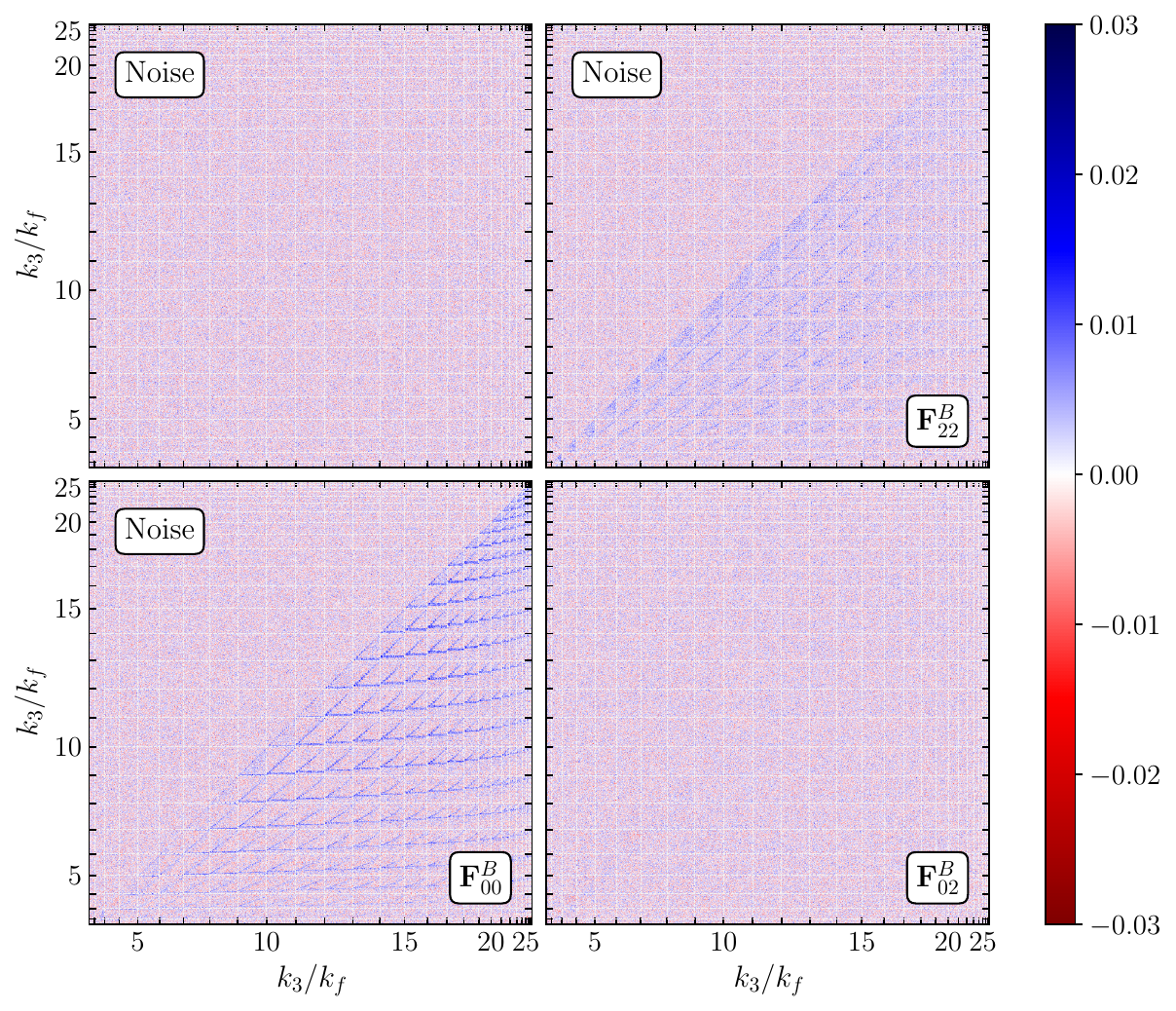}
    \caption{Half-inverse test as defined in equation~\eqref{eq:half-inverse}. The lower-right portion of the plot shows the matrix $\mathbf{F}$. The upper left portion of the plot shows a random matrix with the expected noise level for comparison}
    \label{fig:half-inverse}
\end{figure}

As a second check we perform the ``half-inverse test''~\cite{SlepianEtal2017} to gain more insight on the relevance of all additional contributions not captured by the model. We compute the matrix
\begin{equation}
    \mathbf{F} = \cov_{\mathrm{th}}^{-1/2}\,\hat{\cov}\,\cov_{\mathrm{th}}^{-1/2} - \mathbb{1}\,,
\label{eq:half-inverse}
\end{equation}
where $\hat{\cov}$ is our numerical estimate of the true covariance, measured from $N$ data vector realizations. If the theory covariance is an unbiased model of the true one, $\mathbf{F}$ is a random matrix following a Wishart distribution~\cite{HouEtal2022, Wishart1928} whose variance scales as $1/\sqrt{N}$. We plot the half-inverse matrix for the covariance of the bispectrum in the lower-right portion of figure~\ref{fig:half-inverse}. For comparison, we also plot a Gaussian random matrix with the expected amplitude of the noise in the upper-left portion of the same plot.  The matrix elements where the model deviates the most show a particular structure due to our choice of considering only contributions from  correlations based on sharing the long mode (see \S\ref{sec:non-gaussian-covariance}). Indeed, these are pairs of triangular configurations that share other combination of sides, corresponding to the 8 permutations of equation~\eqref{eq:CBBfull}, so that their analytical covariance is underestimated by $\sim 2\%$ in our simplified model. Overall we see, however, that the model agrees very well with the numerical covariance.

\section{Conclusions}
\label{sec:conclusions}

In this work we provide a test for the approximate model for the bispectrum covariance  proposed in \cite{BiagettiEtal2022}, here extended to redshift-space bispectrum multipoles. This model accounts for the leading non-Gaussian contributions to the covariance expected for squeezed triangular configurations. For these triangles, the non-Gaussian term depending on the product of the power spectrum and trispectrum can be approximated by the product of two bispectra. We consider as well a model for the power spectrum-bispectrum cross-covariance, including all elements where the power spectrum bin coincides with a side of the bispectrum configuration. In this treatment, the main non-Gaussian contributions to the full covariance are written as a function of power spectrum and bispectrum multipoles alone, simplifying their evaluation. Our expressions, obtained in the thin-shell approximation, include as well accurate estimates of the mode-counting factors defined as sums over the Fourier-space grid as integrals in the continuum limit.

We verify the accuracy of our model by comparing it with the numerical covariance obtained from a large suite of 10000 mock halo catalogs. We find that the bispectrum monopole covariance shows significant non-Gaussian contributions for squeezed configurations, of the order of $\sim 10\%$, and that our model correctly recovers these features. We also see that non-Gaussian terms are relatively less important for the covariance of the quadrupole and for the cross covariance between monopole and quadrupole. The same considerations apply for the cross-covariance among power spectrum and bispectrum. The most relevant result, is that our model, despite the approximations, captures all the leading non-Gaussian effects responsible for the off-diagonal structure of the power spectrum and bispectrum covariance matrix. As such it provides a simple analytical expression that can be improved by fitting a few nuisance parameters to a small set of measurements of the data-vector when these are not sufficient for a robust purely numerical estimate of the covariance (see for instance \cite{FumagalliEtal2022}).

There are two main applications of our results. In the first place, the model can replace the simple Gaussian approximation in the analysis of power spectrum and bispectrum measurements from numerical simulations with periodic boundary conditions. It provide an approximate but essentially complete description of the statistical errors involved, including the correlation between the two statistics. In the second place it constitutes one of the ingredients in a more general analytical model for the power spectrum and bispectrum covariance as measured in galaxy redshift surveys, where additional contributions from finite-volume effects should be taken into account. In this case it is possible that even our treatment of squeezed triangular configurations will be affected, due to the characteristic long mode, by window effects. We leave these investigations for a future work. 

\section*{Acknowledgments}

We thank Matteo Biagetti and Kevin Pardede for useful discussions.  L.C. is supported by the STFC Astronomy Theory Consolidated Grant ST/W001020/1 from UK Research $\&$ Innovation. E.S., J.S. and P.M. are partially supported by the INFN INDARK PD51 grant. J.N. is supported by Fondecyt Regular 1211545. E.S. is supported by the INAF Theory grant "NeuMass".

\section*{Additional material}
A Python module providing a numerical implementation of the model for the covariance of the bispectrum (both Gaussian and non-Gaussian) and for the cross-covariance with the power spectrum can be found in the GitLab repository \href{https://gitlab.com/jacopo.salvalaggio/bisque}{\texttt{bisque}}.

\appendix

\section{Integrals in the continuum limit}\label{sec:apx:continuum}

Applying the continuum limit changes sums into integrals:
\begin{gather}
    \sum_{\qv \in k} \simeq \int_{-1}^{1} \de\mu_{\qv} \int_{0}^{2\pi}\de\phi_{\qv}\int_{k-\Delta k/2}^{k+\Delta k/2} \frac{q^2\de q}{k_f^3} \equiv \int_k \frac{\de^3 q}{k_f^3}\;, \\
    \qquad \delta_K(\qv) \simeq k_f^3\,\delta_D(\qv)\,,
\end{gather}
where in the first line we define the shorthand notation for the integral over the $k$-shell: $\mu_{\qv}$ is the cosine of the polar angle $\theta_{\qv}$ between $\qv$ and the LOS $\hat{n}$ and $\phi_{\qv}$ is the azimuthal angle of $\qv$'s rotation around it. $\delta_D$ is the Dirac delta and $k_f$ is the fundamental frequency of the wavevector grid.

First, let us focus on the Gaussian covariance. We can rewrite equation~\eqref{eq:R} in this limit as
\begin{multline}\label{eq:apx:gauss_integral}
    \frac{1}{N_{t}} \sum_{\qv_1\in k_1} \sum_{\qv_2 \in k_2} \sum_{\qv_3 \in k_3} \delta_K(\qv_{123})\,g(\mu_{\qv_1}, \mu_{\qv_2}, \mu_{\qv_3}) \\
    \simeq \frac{1}{N_{t} k_f^6} \int_{k_1} \de^3 q_1 \int_{k_2} \de^3 q_2\int_{k_3} \de^3 q_3 \,\delta_{D}(\qv_{123})\,g(\mu_{\qv_1}, \mu_{\qv_2}, \mu_{\qv_3})\,,
\end{multline}
where, for the sake of brevity, we wrote the argument of the sum as a general angular function $g$. In order to simplify the integral, we first make a change of variables from $\mu_{\qv_2}, \phi_{\qv_2}$ to $\nuq, \xi_{\qv_{12}}$, where the former is the cosine of the angle between $\qv_1$ and $\qv_2$ and the latter is the polar angle of $\qv_2$ around $\qv_1$ as seen in figure \ref{fig:vectors}. In this new basis we have \cite{ScoccimarroCouchmanFrieman1999}
\begin{equation}
    \mu_{\qv_2}(\mu_{\qv_1}, \nuq, \xi_{\qv_{12}}) = \mu_{\qv_1} \nuq - \sqrt{1 - \mu_{\qv_1}^2} \sqrt{1 - \nuq^2} \cos\xi_{\qv_{12}}\,.
\end{equation}
Then, we can simplify the delta function as $\nuq$ only depends on the magnitude of the three sides. In the thin-shell limit:
\begin{equation}\label{eq:apx:nu}
    \nuq^{\star} \simeq \frac{k_3^2 - k_2^2 - k_1^2}{2 k_1 k_2}\,.
\end{equation}
As pointed out in \S\ref{sec:non-gaussian-covariance} this is only applicable to closed and flattened triangles.

The delta function thus becomes:
\begin{equation}
    \del_D (\qv_{123}) = \frac{1}{q_1 q_2 q_3}\del_D(\nuq - \nuq^{\star})\delta_D(\mu_{\qv_3} + \mu_{\qv_{12}})\delta_D(\phi_{\qv_3} + \phi_{\qv_{12}})\,,
\end{equation}
where $\mu_{\qv_{12}}$ and $\phi_{\qv_{12}}$ are the angular polar coordinates of $\qv_{12} \equiv \qv_1 + \qv_2$. Plugging this into equation~\eqref{eq:apx:gauss_integral} we get:
\begin{multline}\label{eq:apx:fullPPP}
    \frac{1}{N_{t}} \sum_{\qv_1\in k_1} \sum_{\qv_2 \in k_2} \sum_{\qv_3 \in k_3} \delta_K(\qv_{123})\,g(\mu_{\qv_1}, \mu_{\qv_2}, \mu_{\qv_3}) \\
    \simeq \frac{1}{4\pi} \int\de\mu_{\qv_1}\int\de\xi_{\qv_{12}}\,g\left(\mu_{\qv_1}, \mu_{\qv_2}(\mu_{\qv_1}, \xi_{\qv_{12}}), -\mu_{\qv_{12}}(\mu_{\qv_1}, \xi_{\qv_{12}})\right)\,,
\end{multline}
where the dependence on $\nuq$ was dropped as for any given triangle it is a fixed quantity, as shown in equation~\eqref{eq:apx:nu}.
A Python function containing the exact solution for $\cov_{\ell_1 \ell_2}^{B(PPP)}$ (equation \eqref{eq:C_PPP}) for $\ell_1, \ell_2 = 0, 2$ is provided in the module \texttt{gaussian.py} of \texttt{bisque}.

Moving on to the non-Gaussian contributions to the bispectrum covariance, we will provide here a detailed expression for the $I$ term in equation~\eqref{eq:finalBB}. The first steps are analogous to what was shown earlier. Carrying out all straightforward integrals one is left with
\begin{align}
    I_{\ell_1 \ell_2 \ell_3 \ell_4}(t_i, t_j) = \frac{1}{8\pi^2}\int & \,\de\mu_{\qv_1}\de\xi_{\qv_{12}}\de\mu_{\pv_1}\de\xi_{\pv_{12}}\del_D(\mu_{\qv_{12}} - \mu_{\pv_{12}}) \nonumber\\
    & \times \mathcal{L}_{\ell_1}(\mu_{\qv_1})\mathcal{L}_{\ell_2}(\mu_{\pv_1})\mathcal{L}_{\ell_3}(\mu_{\qv_1})\mathcal{L}_{\ell_4}(\mu_{\pv_1})\,.
\end{align}
We will proceed by solving the Dirac delta argument for $\xi_{\pv_{12}}$. There are two roots in the $[0, 2\pi)$ interval, $\xi_{\pv_{12}}^{\pm}$:
\begin{equation}\label{eq:xippm}
	\cos\xi_{\pv_{12}}^{\pm}  = \frac{(k_{1,i} + k_{2,i}\,\mu_{\qv_{12}})\mu_{\qv_1} - (k_{1,j} + k_{2,j}\,\mu_{\pv_{12}})\mu_{\pv_1} - k_{2,i}\sqrt{1 - \mu_{\qv_1}^2}\sqrt{1 - \mu_{\qv_{12}}^2}\cos\xi_{\qv_{12}}}{k_{2,j}\sqrt{1 - \mu_{\pv_1}^2}\sqrt{1 - \mu_{\pv_{12}}^2}} \equiv \zeta_{\pv_{12}}\,.
\end{equation}
After integrating out the delta we get
\begin{equation}\label{eq:apx:fullBB}
    I_{\ell_1 \ell_2 \ell_3 \ell_4}(t_i, t_j) = \frac{1}{2\pi^2} \int \de\mu_{\qv_1}\de\xi_{\qv_{12}}\int_{\mathcal{D}}\de\mu_{\pv_1} \frac{k_{3,j} \mathcal{L}_{\ell_1}(\mu_{\qv_1})\mathcal{L}_{\ell_2}(\mu_{\pv_1})\mathcal{L}_{\ell_3}(\mu_{\qv_1})\mathcal{L}_{\ell_4}(\mu_{\pv_1})}{k_{2,j}\sqrt{1-\nu_{\pv_{12}}^{\star\,2}}\,\sqrt{1-\mu_{\pv_1}^2}\,\sqrt{1-\zeta_{\pv_{12}}^2(\mu_{\qv_1}, \xi_{\qv_{12}}, \mu_{\pv_1})}}\,,
\end{equation}
where $\mathcal{D}$ is the set of values of $\mu_{\pv_1}$ where the integral is well-defined, i.e. where $\zeta_{\pv_{12}}^2 < 1$. A Python function to solve equation~\eqref{eq:apx:fullBB} numerically can be found in the module \texttt{non\_gaussian.py} of \texttt{bisque}.

Finally, we show the thin-shell expression for the cross-covariance factor $\Tilde{I}$. The steps are similar to the ones shown above. After performing some angular integrals and integrals over the magnitudes $q_1$, $q_2$, $q_3$, and simplifying a factor of $N_{t}$ the quantity defined in equation~\eqref{eq:tildeI} can be written as
\begin{align}\label{eq:apx:fullPB}
    \Tilde{I}_{\ell_1\ell_2\ell_3\ell_4}(k_i, t_j) = \frac{1}{2 N_{k_i}} \int &\,\de\mu_{\qv_1}  \mathcal{L}_{\ell_2}(\mu_{\qv_1})\mathcal{L}_{\ell_4}(\mu_{\qv_1}) \nonumber \\
    &\times\left[\,\delta^K_{k_i,k_{1,j}} \mathcal{L}_{\ell_1}(\mu_{\qv_1})\mathcal{L}_{\ell_3}(\mu_{\qv_1}) \right.\nonumber \\ 
    & \left.\quad\,\; + \delta^K_{k_i,k_{2,j}} \mathcal{L}_{\ell_1}(\mu_{\qv_2}(\mu_{\qv_1}, \xi_{\qv_{12}}))\mathcal{L}_{\ell_3}(\mu_{\qv_2}(\mu_{\qv_1}, \xi_{\qv_{12}})) \right.\nonumber \\
    & \left. \quad\,\; + \delta^K_{k_i,k_{3,j}} \mathcal{L}_{\ell_1}(-\mu_{\qv_{12}}(\mu_{\qv_1}, \xi_{\qv_{12}}))\mathcal{L}_{\ell_3}(-\mu_{\qv_{12}}(\mu_{\qv_1}, \xi_{\qv_{12}})) \right]\,. 
\end{align}
An exact solution of equation~\eqref{eq:apx:fullPB} for $\ell_1, \ell_2 = 0, 2$ is coded in the module \texttt{cross.py} of \texttt{bisque}.

\bibliographystyle{JHEP}
\bibliography{cosmologia}

\end{document}